%% file: partsmoothMH.tex
\documentclass[11pt]{article}

\usepackage{amsmath,amssymb,amsfonts,ifthen,latexsym,mathrsfs,enumerate,eucal,bbm,subfigure,verbatim}
\usepackage{graphicx,color,framed,stmaryrd}
\usepackage{graphicx,color,framed,gensymb,upgreek,twoopt}
\usepackage{amsbsy}
\usepackage[numbers]{natbib}
\usepackage{algorithmicx}
\usepackage{algpseudocode}
\usepackage[numbered]{algo}

\usepackage{graphics}
\usepackage[dvips]{epsfig}

\numberwithin{equation}{section}
\input{partsmoothMH_def}

\begin{document}

\title{Metropolising forward particle filtering backward sampling
and Rao-Blackwellisation of Metropolised particle smoothers}

\author{Jimmy Olsson \and Tobias Ryd\'en}

\maketitle

\begin{abstract}
\input{partsmoothMH_abstract}
\end{abstract}

\setcounter{equation}{0}
\input{partsmoothMH_intro}

\section{Preliminaries}
\label{section:prel}
\input{partsmoothMH_preliminaries}

\section{Algorithms}
\label{section:alg}

\subsection{Particle smoothing}
\label{section:PF}
\input{partsmoothMH_PF}

\subsection{FFBS-based independent Metropolis-Hastings sampler}
\label{section:PIMH}
\input{partsmoothMH_PIMH}

\subsection{Rao-Blackwellisation and multiple trajectories}
\label{sec:RB}
\input{partsmoothMH_RB}

\subsection{Including a parameter}
\label{section:PMMH}
\input{partsmoothMH_parameter}

\section{Example}
\label{section:example}
\input{partsmoothMH_examp}

\appendix

\section{Proofs}
\label{section:proofs}
\input{partsmoothMH_proofs}

\bibliographystyle{chicago}
\bibliography{motherofallbibs}

\end{document}

%% file: partsmoothMH_def.tex
\newcommand{\adj}[2]{\vartheta_{#1}^{#2}}
\newcommand{\adjfunc}[4][]
{\ifthenelse{\equal{#1}{}}{\ifthenelse{\equal{#4}{}}{\vartheta_{#2}}{\vartheta_{#2}(#4)}}
{\ifthenelse{\equal{#1}{smooth}}{\ifthenelse{\equal{#4}{}}{\tilde{\vartheta}_{#2}}{\tilde{\vartheta}_{#2}(#4)}}
{\ifthenelse{\equal{#1}{fully}}{\ifthenelse{\equal{#4}{}}{\vartheta^\star_{#2}}{\vartheta^\star_{#2}(#4)}}{\mathrm{erreur}}}}
}

\newcommand{\ap}[2][]{\ifthenelse{\equal{#1}{}}{\alpha_{#2}}{\alpha_{#2}^N}}

\newcommand{\auxMHprop}[1]{k^N_{#1}}
\newcommand{\auxMHtarg}[1]{\pi_{#1}^N}
\newcommand{\auxprop}[1]{\Pi_{#1}^N}
\newcommand{\backind}[1]{J_{#1}}

\newcommand{\boralg}{\mathcal{B}}
\newcommand{\chunk}[4][]%
{\ifthenelse{\equal{#1}{}}{\ensuremath{{#2}_{#3:#4}}}{\ensuremath{#2^#1}_{#3:#4}}}

\newcommand{\cond}{\,|\,}

\newcommand{\E}{\mathbb{E}}
\newcommand{\eg}{e.g.}
\newcommand{\epart}[2]{\ensuremath{\xi_{#1}^{#2}}}
\newcommand{\epartset}[2][]{\ifthenelse{\equal{#1}{}}{{\ensuremath{\boldsymbol{\xi}_{#2}}}}{\ensuremath{\boldsymbol{\xi}_{#2}}}}
\newcommand{\eqdef}{\ensuremath{\stackrel{\mathrm{def}}{=}}}
\newcommand{\eqsp}{\;}
\newcommand{\esssup}[2][]%
{\ifthenelse{\equal{#1}{}}{\left| #2 \right|_\infty}{\left| #2 \right|^2_{\infty}}}
\newcommand{\ewght}[2]{\ensuremath{\omega_{#1}^{#2}}}
\newcommand{\ewghtset}[1]{\ensuremath{\boldsymbol{\omega}_{#1}}}
\newcommand{\ewghtfunc}[1]{\ensuremath{\omega_{#1}}}

\newcommand{\filt}[2][]%
{
\ifthenelse{\equal{#1}{}}{\ensuremath{\phi_{#2}}}%
{\ifthenelse{\equal{#1}{hat}}{\ensuremath{\phi^{N}_{#2}}}
{\ifthenelse{\equal{#1}{tilde}}{\ensuremath{\tilde{\phi}^{N}_{#2}}}
{\ifthenelse{\equal{#1}{tar}}{\ensuremath{\phi^{N,\mathrm{t}}_{#2}}}
{\ifthenelse{\equal{#1}{aux}}{\ensuremath{\phi^{N,\mathrm{aux}}_{#2}}}
}
}
}
}
}

\newcommand{\genind}[2]{G_{#1}^{#2}}

\newcommand{\ie}{i.e.}

\newcommand{\ind}[2]{I_{#1}^{#2}}
\newcommand{\indset}[2][]{\ifthenelse{\equal{#1}{}}{\ensuremath{\mathbf{i}_{#2}}}{\ensuremath{\mathbf{I}_{#2}}}}
\newcommand{\indic}{\mathbbm{1}}
\newcommand{\jointpartlaw}[1]{\psi^N_{#1}}
\newcommand{\M}{\ensuremath{Q}}
\newcommand{\Mdens}{\ensuremath{q}}

\newcommand{\mcbf}[2][]{%
\ifthenelse{\equal{#1}{}}{\overline{\mathcal{F}}_{#2}}{\overline{\mathcal{F}}_{#2}^{(#1)}}%
}
\newcommand{\MHch}[2]{X_{#1}^{(#2)}}
\newcommand{\MHprop}[2][]%
{
\ifthenelse{\equal{#1}{}}{\ensuremath{k_{#2}}}{\ensuremath{k^N_{#2}}}%
}

\newcommand{\parind}{}
\newcommand{\partdraw}[2]{\ensuremath{X_{#1}^\ast}}

\newcommand{\plim}{\ensuremath{\stackrel{\mathbb{P}}{\longrightarrow}}}
\newcommand{\post}[3][]%
{
\ifthenelse{\equal{#1}{}}{\ensuremath{\phi_{#2|#3}}}%
{\ifthenelse{\equal{#1}{hat}}{\ensuremath{\phi^{N}_{#2|#3}}}
{\ifthenelse{\equal{#1}{tilde}}{\ensuremath{\tilde{\phi}^{N}_{#2|#3}}}
{\ifthenelse{\equal{#1}{tar}}{\ensuremath{\phi^{N,\mathrm{t}}_{#2|#3}}}}
}
}
}

\newcommand{\prob}{\mathbb{P}}

\newcommand{\proposal}[1]{R_{#1}}
\newcommand{\propdens}[1]{r_{#1}}
\newcommand{\R}{\mathbb{R}}
\newcommand{\refm}{\lambda}

\newcommand{\revM}[1]{\overleftarrow{\M}_{#1}}
\newcommand{\rmd}{\ensuremath{d}}
\newcommand{\supnorm}[1]{\ensuremath{\left\|#1\right\|_{\infty}}}

\newcommand{\ub}{q_+}

\newcommand{\oscnorm}[2][]%
{\ifthenelse{\equal{#1}{}}{\ensuremath{\operatorname{osc}\left(#2\right)}}{\ensuremath{\operatorname{osc}^{#1}\!\left(#2\right)}}}
\newcommand{\Xinit}{\ensuremath{\rho}}
\newcommand{\XinitIS}[2][]
{\ifthenelse{\equal{#1}{}}{\ensuremath{\rho_{#2}}}{\ensuremath{\check{\rho}_{#2}}}}
\newcommand{\Xsigma}[1][]%
{\ifthenelse{\equal{#1}{}}{\ensuremath{\mathcal{B}(\Xset)}}{\ensuremath{\mathcal{B}(\Xset^{#1})}}}
\newcommand{\Xset}{\ensuremath{\mathcal{X}}}
\newcommand{\Yrefm}{\nu}
\newcommand{\Yset}{\ensuremath{\mathcal{Y}}}
\newcommand{\Ysigma}[1][]%
{\ifthenelse{\equal{#1}{}}{\ensuremath{\mathcal{B}(\Yset)}}{\ensuremath{\mathcal{B}(\Yset^{#1})}}}
\newcommand{\wrt}{with respect to}
\newcommand{\Xchain}{\mathbf{X}}

\newcommand{\Yproc}{\mathbf{Y}}

\renewcommand{\mid}{\, | \,}

\newtheorem{thm}{Theorem}

\newtheorem{cor}[thm]{Corollary}

\newtheorem{assum}{A}

%% file: partsmoothMH_abstract.tex
Smoothing in state-space models amounts to computing the
conditional distribution of the latent state trajectory,
given observations, or expectations of functionals of
the state trajectory with respect to this distributions.
For models that are not linear Gaussian or possess finite
state space, smoothing distributions are in general
infeasible to compute as they involve intergrals over
a space of dimensionality at least equal to the number of
observations. Recent years have seen an increased
interest in Monte Carlo-based methods for smoothing,
often involving particle filters. One such method is to
approximate filter distributions with a particle filter,
and then to simulate backwards on the trellis of particles
using a backward kernel. We show that by supplementing this
procedure with a Metropolis-Hastings step deciding whether to
accept a proposed trajectory or not, one obtains a Markov chain
Monte Carlo scheme whose stationary distribution is the exact
smoothing distribution. We also show that in this procedure,
backward sampling can be replaced by backward smoothing,
which effectively means averaging over all possible trajectories.
In an example we compare these approaches to
a similar one recently proposed by
Andrieu, Doucet and Holenstein,
and show that the new methods can be more efficient in terms of precision
(inverse variance) per computation time.

%% file: partsmoothMH_intro.tex
\section{Introduction}

The topic of the present paper is computation of smoothed expectations
of functionals of the state process in state-space models,
\ie\ conditional expectations of such functionals given data.
To make this discussion more precise, let $(\Xchain,\Yproc)$
be a \textit{state-space model}, where $\Yproc = (Y_k)_{k=0}^n$ is the
observable (output) process and $\Xchain = (X_k)_{k=0}^n$
is the the latent (unobserved) Markov chain. The relation
between the two is such that given $\Xchain$, the $Y_k$'s are
conditionally independent with the conditional distribution of a
particular $Y_k$ depending on the corresponding $X_k$ only.
If the state space of $\Xchain$ is finite we use the term
\textit{hidden Markov model} (HMM). Our interest thus lies in
computing conditional expectations of the form
$\E[h(X_{0:n})\mid y_{0:n}]$ for a a real-valued
functional $h$ of one, several, or all of the latent variables.
Here $X_{0:k} = (X_0, X_1, \ldots, X_k)$ etc.;
this form will be our generic notation for vectors.

Conditional expectations as above are often interesting and relevant
in their own respect, with \eg\ $\E[X_k\mid Y_{0:n}]$,
$\E[X_k^2\mid Y_{0:n}]$ and
$\prob(X_k\geq x\mid Y_{0:n}) = \E[\mathbbm{1}(X_k \geq x) \mid Y_{0:n}]$,
where $\mathbbm{1}(\cdot)$ denotes some indicator function, providing useful inferential summaries of the latent states.
Another very common use of such expectations is however for
inference on model parameters through the EM algorithm.
Indeed, assume that the distribution of $(\Xchain,\Yproc)$ depends
on some model parameter (vector) $\theta$. Then in the E-step of
the EM algorithm, it is typical that conditional expectations
with functionals like $h(x_{0:n})=\sum_{k=0}^n x_k$,
$h(x_{0:n})=\sum_{k=0}^n x_k^2$,
$h(x_{0:n})=\sum_{k=1}^n x_{k-1}x_k$ etc.\ appear, in particular
if the joint distribution of $(\Xchain,\Yproc)$ belongs to an
exponential family of distributions.
For HMMs, the functional $h(x_{0:n})=\sum_{k=1}^n \mathbbm{1}(x_{k-1}=i,x_k=j)$
is used to re-estimate the transition probability from state $i$
to state $j$.
Unfortunately, exact numeric computation of the conditional distribution
of $X_k$, or that of a sequence of $X$'s, given $Y_{0:n}$, is possible
essentially only in two cases. Firstly for HMMs, for which the
forward-backward algorithm \citep{baum:petrie:soules:weiss:1970}
provides the solution,
and secondly for linear Gaussian state-space models, for which
conditional distributions are Gaussian and the Kalman smoothing
recursions provide their conditional means and (co)variances
\citep[e.g.][Chapter~4]{durbin:koopman:2001}.
For other models, \ie\ models with continuous state space and
non-linear and/or non-Gaussian dynamics and/or output characteristics,
there are no exact numerical methods available and one is confined to
using approximations. Traditional approaches include Kalman filtering
and smoothing techniques based on linearisation of the system
dynamics and output characteristics, such as the extended Kalman filter,
but following the impact of Markov chain
Monte Carlo (MCMC) methods in general, during the last 10--15 years
there has been a dramatic increase in the interest in and use of
simulation-based methods to approximate conditional expectations
given data. When such methods are used to approximate expectations
appearing in the E-step of the EM algorithm, one often talks about
\textit{Monte Carlo EM} (MCEM) algorithms.

For a model as above it holds that the conditional distribution of
$\Xchain$ given $\Yproc$ is that of a time-varying Markov chain. This
fact lies behind the existence of the forward-backward
algorithm for HMMs, and is also the cornerstone of any algorithm that
simulates $\Xchain$ conditionally on $\Yproc$.
To simulate $\Xchain$ given $\Yproc$ and a fixed set of parameters
$\theta$, there are essentially two different approaches. The first one,
often referred to as \textit{local updating},
is to run an MCMC algorithm that updates one $X_k$ at the time,
given $\Yproc$ \textit{and} $X_{-k}=(X_j)_{0\leq j\leq n, j\neq k}$.
Because of the model structure, the only variables appearing in this
conditional distribution are $Y_k$, $X_{k-1}$ (unless $k=0$) and
$X_{k+1}$ (unless $k=n$). By varying $k$ one obtains an MCMC
algorithm whose stationary distribution is that of $\Xchain$ given $\Yproc$
\citep[see \eg][]{robert:celeux:diebolt:1993}.
The other approach to simulating $\Xchain$ given $\Yproc$ is
simply to simulate a full trajectory from the conditional distribution
in question. This can be done either by \textit{forward filtering-backward
sampling} (FFBS), or by \textit{backward recursion-forward sampling}.
These names stem from the two blocks of the forward-backward algorithm
for HMMs, replacing either of them by simulation. In this paper we
focus on the former approach. After recursively computing the filter,
\ie\ the conditional distribution of $X_k$ given $Y_{0:k}$, for
$k = 0, 1, \ldots, n$, forward filtering-backward sampling first simulates
$X_n$ from the filter distribution at time $n$ and then recursively
simulates, for $k = n-1, n-2, \ldots, 0$, $X_k$ from the conditional
distribution of $X_k$ given $X_{k+1}$ and $Y_{0:k}$.

Comparing the two approaches, the advantage of local updating is its
simplicity as it only involves simulation from univariate (conditional)
distributions. Its disadvantage is that a significant burn-in period 
may be required to remove bias, and that mixing can be slow so that 
many MCMC iterations are required to make sure that sample means 
approximate the corresponding conditional expectations with required accuracy.
For FFBS on the other hand, one must compute the filter distribution.
This is easily done for HMMs \citep[e.g.][used FFBS for HMMs]{chib:1996},
but for continuous state spaces the filter distributions are in general
not available. The foremost advantage of FFBS is that the simulated
replications of $\Xchain\mid\Yproc$ are independent.

To approximate filter distributions in models with continuous state space,
a class of methods known as \emph{particle filters}, or
\emph{sequential Monte Carlo} (SMC) methods, has received considerable
attention during the last 10--15 years; see \eg\
\cite{fearnhead:1998,doucet:defreitas:gordon:2001,cappe:moulines:ryden:2005}
for introductions to such methods, and \eg\
\cite{andrieu:doucet:sumeetpal:tadic:2004,gustafsson:2010,
schon:gustafsson:karlsson:2011} for surveys and applications.

Particle filters approximate the filter distribution at time $k$
by a discrete distribution
$\sum_{i = 1}^N \ewght{k}{i} \delta_{\epart{k}{i}}$, for which the
locations $\epart{k}{i}$, the so-called \emph{particles}, and the
non-negative weights $\ewght{k}{i}$ evolve randomly and recursively in
time. Using such an approximation one can thus recursively compute
approximations to the filter distributions, and then use them
to simulate a state trajectory backwards. The distribution of this
trajectory will then only approximately be that of $\Xchain\mid\Yproc$.
The idea to use particle filters for approximate backward
sampling first appeared, to the best of our knowledge, in
\cite{doucet:godsill:andrieu:2000}. The paper 
\cite{douc:garivier:moulines:olsson:2009} 
provides theory (see e.g.\ Theorem~5 and Corollary~6 therein) that
supports the validity of this approach such as consistency results
ensuring that, as the number of particle increases, the distribution
of a trajectory sampled using the particle filter converges to the
true smoothing distribution.

A recent paper, \cite{andrieu:doucet:holenstein:2010}, 
devised a
method related to FFBS but that removes bias entirely by adding a
Metropolis-Hastings (M-H) step. The approach is described in detail below,
but in short it involves running a particle filter and then
selecting a state trajectory not by backward sampling, but by
sampling one of the particles at the final time-point $n$
according to its importance weight, and then tracing the history
of this particle back to the first time-point. The M-H step
is constructed so that the stationary distribution of the sampled
trajectory is indeed the distribution of $\Xchain\mid\Yproc$.
A well-known problem of particle filters is however that as the
filter recursion proceeds beyond a given time-point $k$, after
a while only a few of the particles that existed at $k$ will
have survived the step-wise selection process. This implies
that the genealogical tree of the particle filter provides a poor
approximation to the smoothing distribution at time-points
just a bit prior to current time. FFBS does not suffer from this kind of
degeneration, and in the present paper we show how an M-H step
can be applied to remove bias also from particle FFBS.
We also show how the approach can be Rao-Blackwellised, by which we
mean that sampling of trajectories is replaced by the corresponding
expectation, which is the backward smoothing recursion.
As a compromise between single trajectory sampling and smoothing
one may also simulate a small number of trajectories from each
set of particles. We analyse the various approaches from a
variance/cost perspective, and show that backward simulation and
smoothing can be notablty more efficient than sampling from the
genealogical tree.

We started the work on the material presented in this paper
when we during the writing of the manuscript
\cite{olsson:ryden:stjernqvist:2010}, on approximate data augmentation
MCMC schemes using particle FFBS,
became aware of a preprint version of 
\cite{andrieu:doucet:holenstein:2010}. 
Later, in the discussion part
following the published version of that paper (p.~306--307), we
found that Nick Whitley (University of Bristol) had been
thinking along similar lines. Therefore we would like to point out
some features of our paper that are not found in Whiteley's comment,
nor in the paper 
\cite{andrieu:doucet:holenstein:2010} 
itself.
One such feature is that we allow for general auxiliary particle
filters in the MCMC sampler, and another one is the multiple
trajectory sampling idea that compromises between single trajectory
sampling and backward smoothing, as well the variance/cost analysis of
this and other approaches.

%% file: partsmoothMH_preliminaries.tex
In this section we sharpen the notation and introduce some basic
concepts that will be used throughout the paper.
We assume that all random variables are defined on a common
probability space $(\Omega, \mathcal{F}, \prob)$. The state space of
$\Xchain$ is denoted by $\Xset$, and by $\Yset$ we denote the space
in which $\Yproc$ takes its values. We suppose that both of these
spaces are Polish and write $\Xsigma$ and $\Ysigma$ respectively
for the corresponding Borel $\sigma$-fields.
The transition kernel and initial distribution of $\Xchain$ are denoted by $\M$ and $\Xinit$, respectively,
and we assume that the transition kernel $\M$ admits a density $\Mdens$ w.r.t.
some fixed reference measure measure $\refm$ on $\Xset$, in the sense that
$$
    \M(x, A) = \int \mathbbm{1}_A(x') \Mdens(x, x') \, \refm(\rmd x')
$$
for all $A \in \Xsigma$ and $x \in \Xset$.
We also assume that the conditional distribution of $Y_k$ given
$X_k = x$ has a density $g(x, \cdot)$ (the \textit{emission density})
w.r.t\ some reference measure $\Yrefm$. In most applications $\Xset$
and $\Yset$ are products of $\R$, and $\refm$ and $\Yrefm$ are
Lebesgue measures. Here we have tacitly
assumed that neither $\M$ nor $g$ depends on
time $k$, but the extension to time-varying
systems is immediate.

We will throughout the paper assume that we are given a fixed record $y_{0:n}$
of arbitrary but fixed observations and our main target is to produce
samples from the joint posterior distribution $\post{r:s}{n}(A \parind) \eqdef
\prob(X_{r:s}\in A\cond Y_{0:n} = y_{0:n}\parind)$,
$A \in \boralg(\Xset)^{\varotimes (s - r + 1)}$, of a record
of states given the observations. The special cases $\filt{k}
\eqdef \post{k:k}{k}$ and $\post{0:n}{n}$ will be
referred to as the \emph{filter} and \emph{joint smoothing}
distributions, respectively. Since the model is fully dominated, each joint
smoothing distribution $\post{0:k}{k}$ has a density (denoted by the same symbol) w.r.t.
products of $\refm$ and $\Yrefm$. This density is proportional to
$\Xinit(x_0) g_0(x_0) \prod_{\ell = 1}^k g_\ell(x_\ell) \Mdens(x_{\ell - 1}, x_\ell)$
and we denote by $Z_k$ the normalising constant. Since the
observations are fixed we will keep the dependence of any
quantity on these implicit and introduce the
short-hand notation $g_k(x \parind) \eqdef g(x, y_k \parind)$
for $x \in \Xset$.

It is easily shown that the joint
smoothing distributions $(\post{0:k}{k})_{k = 0}^n$ satisfy the well
known \emph{forward smoothing recursion}
\begin{equation} \label{eq:smoothing:recursion}
    \post{0:k}{k}(A \parind)
    = \frac{\iint \indic_A(x_{0:k}) g_k(x_k \parind) \,
    \M(x_{k - 1}, \rmd x_k \parind) \,
    \post{0:k - 1}{k - 1}(\rmd x_{0:k - 1} \parind)}
    {\iint g_k(x_k \parind) \, \M(x_{k - 1}, \rmd x_k \parind) \,
    \post{0:k - 1}{k - 1}(\rmd x_{0:k - 1} \parind)} \eqsp,
\end{equation}
implying the analogous recursion
\begin{equation} \label{eq:filter:recursion}
    \filt{k}(A \parind) = \frac{\iint \indic_A(x_k) g_k(x_k \parind) \,
    \M(x_{k - 1}, \rmd x_k \parind) \, \filt{k - 1}(\rmd x_{k - 1} \parind)}
    {\iint g_k(x_k \parind) \, \M(x_{k - 1}, \rmd x_k \parind) \,
    \filt{k - 1}(\rmd x_{k - 1} \parind)}
\end{equation}
for the filter distributions. Conversely, the joint smoothing distributions
may be retrieved from the filter distribution flow using the so-called \emph{backward
decomposition} of the smoothing measure. Indeed, let, for $A \in \Xsigma$,
\begin{equation} \label{eq:backward:kernel:density:form}
    \revM{\mu}(x', A \parind) = \frac{\int \indic_A(x) \, \Mdens(x, x' \parind) \,
    \mu(\rmd x)}{\int \Mdens(u, x' \parind) \, \mu(\rmd u)}
\end{equation}
be the \emph{reverse kernel} associated with $\M$ and $\mu$, where $\mu$ is
a probability measure on $\Xset$. In particular, by letting $\mu$ be some marginal
distribution of $\Xchain$ we obtain the transition kernel of $\Xchain$
when evolving in reverse time. Using \eqref{eq:backward:kernel:density:form},
the joint smoothing distribution $\post{0:n}{n}$ may be expressed as
\begin{equation} \label{eq:smoothing:backw:decomposition}
    \post{0:n}{n}(A \parind) = \idotsint \indic_A(x_{0:n}) \,
    \filt{n}(\rmd x_n \parind) \, \prod_{k = 0}^{n - 1}
    \revM{\filt{k}}(x_{k + 1}, \rmd x_k \parind)
\end{equation}
for $A \in \Xsigma^{\varotimes (n + 1)}$
\citep[Corollary~3.3.8]{cappe:moulines:ryden:2005}. Here $( \revM{\filt{k}} )_{k = 0}^{n - 1}$
are the so-called \emph{backward kernels} describing transitions of $\Xchain$ when evolving
backwards in time and \emph{conditionally} on the given observations. Consequently, a draw $X_{0:n}$ from $\post{0:n}{n}$ may be produced by first computing recursively, using \eqref{eq:filter:recursion}, the filter distributions $(\filt{k})_{k = 0}^n$ (the forward filtering pass), simulating $X_n$ from $\filt{n}$, and then simulating, recursively for $k = n - 1, n - 2, \ldots, 0$, $X_k$ from $\revM{\filt{k}}(X_{k + 1},\cdot\parind)$ (the backward simulation pass). This is the mentioned FFBS algorithm.

As stressed in the introduction, joint smoothing distributions can be expressed on closed form only for a very few models. The same applies for any marginals of the same, including the filter distributions, and thus the decomposition \eqref{eq:smoothing:backw:decomposition} appears, at a first glance, to be of academic interest only. However, while there is a well-established difficulty of applying SMC methods directly to the smoothing recursion \eqref{eq:smoothing:recursion} (as resampling systematically the particle trajectories decreases rapidly the number of distinct particle coordinates at early time steps; see Section~\ref{section:PF}), SMC methods may be efficiently used for approximating the filter distributions. Hence, by following \cite{doucet:godsill:andrieu:2000} and replacing the filter distributions in \eqref{eq:smoothing:backw:decomposition} by particle filter estimates, we obtain an approximation of the joint smoothing distribution that is not at all effected by the degeneracy of the genealogical particle tree. This issue will be discussed further in Section~\ref{section:PF}.

%% file: partsmoothMH_PF.tex
A particle filter approximates the filter distribution
$\filt{k}$ at time $k$ by a weighted empirical measure
\begin{equation}\label{eq:partapprox}
    \filt[hat]{k}(\rmd x \parind) \eqdef \sum_{i = 1}^N \frac{\ewght{k}{i}}
    {\sum_{\ell = 1}^N \ewght{k}{\ell}} \delta_{\epart{k}{i}}(\rmd x) \eqsp,
\end{equation}
where $(\epart{k}{i}, \ewght{k}{i})_{i = 1}^N$ is
a weighted finite sample of so-called particles (the $\epart{k}{i}$'s) with associated importance weights (the $\ewght{k}{i}$'s) and $\delta_{\epart{}{}}$ denotes a unit point mass at $\epart{}{}$. We remark that the weights $(\ewght{k}{i})_{i=1}^N$ are not
normalised, \ie\ required to sum to unity, which motivates the
self-normalisation in \eqref{eq:partapprox}.

Based on an approximation as above at time $k$, an approximation of
$\filt{k+1}$ can be obtained in different ways; however, two specific operations are common to all SMC algorithms: \textit{selection}, which amounts
to dropping particles that have small importance weights and
duplicating particles with larger weights, and \textit{mutation},
which amounts to randomly moving the particles in the state space $\Xset$.
The approach we describe below is called the
\emph{auxiliary particle filter} \citep{pitt:shephard:1999}.

Given the ancestor sample $(\epart{k}{i}, \ewght{k}{i})_{i = 1}^N$, one
iteration of the auxiliary particle filter involves sampling
the auxiliary distribution
\begin{multline*} \label{eq:auxiliary:target}
     \Phi^N_{k + 1}(\{ i \} \times A \parind)
     \\ \eqdef \frac{\ewght{k}{i} \int g_{k + 1}(x \parind) \,
     \M(\epart{k}{i}, \rmd x \parind)}{\sum_{\ell = 1}^N \ewght{k}{\ell}
     \int g_{k + 1}(x \parind) \, \M(\epart{k}{\ell}, \rmd x \parind)}
     \left(\frac{\int \indic_A(x) g_{k + 1}(x \parind) \,
     \M(\epart{k}{i}, \rmd x \parind)}{\int g_{k + 1}(x \parind) \,
     \M(\epart{k}{i}, \rmd x \parind)} \right)
\end{multline*}
on the product space $\Xset \times \{1, \ldots, N\}$, using some
proposal distribution
\begin{equation*} \label{eq:auxiliary:proposal}
    \auxprop{k + 1}(\{ i \} \times A \parind) \eqdef \frac{\ewght{k}{i}
    \adj{k}{i}}{\sum_{\ell = 1}^N \ewght{k}{\ell}
    \adj{k}{\ell}}
    \proposal{k}(\epart{k}{i}, A \parind)
\end{equation*}
where $\proposal{k}$ is a proposal kernel on
$\Xset$ and $(\adj{k}{i})_{i = 1}^N$ is a set of adjustment
multiplier weights. As a motivation for this we note that $\sum_{i = 1}^N \auxprop{k + 1}(\{ i \} \times A \parind)$ is the mixture distribution obtained by simply plugging the weighted empirical measure \eqref{eq:partapprox} into the filtering recursion \eqref{eq:filter:recursion}; thus, by simulating a set of particle positions and indices from \eqref{eq:auxiliary:proposal} and discarding the latter, a sample of particles approximating $\filt{k + 1}$ is obtained. This procedure may then be repeated recursively as new observations become available in order to obtain weighted particle samples approximating the filter distributions at all time points. We will throughout this paper assume that the
adjustment multiplier weights are generated from the ancestor sample
according to $\adj{k}{i} \eqdef \vartheta_k(\epart{k}{i} \parind)$,
where $\vartheta_k : \Xset \rightarrow \R^+$ is weight function.
In addition we will assume that the proposal kernel has a transition density
$\propdens{k} : \Xset^2 \rightarrow \R^+$ with respect to $\refm$.
The latter implies that also $\auxprop{k + 1}$ has a density,
which we denote by the same symbol, on $\{ 1, \ldots, N \} \times \Xset$.
In practice a draw from $\Pi^N_{k + 1}$ is produced by first drawing
an index $\ind{}{} = i$ with probability proportional to
$\ewght{k}{i} \adj{k}{i}$ and then simulating a new particle location
$\epart{}{}$ from the measure $\proposal{k}(\epart{k}{I}, \rmd x \parind)$.
Each of the draws $(\epart{k + 1}{i}, \ind{k + 1}{i})_{i = 1}^N$
from $\Pi^N_{k + 1}$ is assigned the importance weight
\begin{equation*}
    \ewght{k + 1}{i} \eqdef \ewghtfunc{k + 1}
    ( \epart{k}{\ind{k + 1}{i}}, \epart{k + 1}{i} \parind )
    \propto \frac{\rmd \Phi^N_{k + 1}}{\rmd \Pi^N_{k + 1}}
    ( \epart{k + 1}{i}, \ind{k + 1}{i} \parind ) \eqsp,
\end{equation*}
where $\ewghtfunc{k + 1}(\cdot \parind) : \Xset^2 \rightarrow \R^+$ is the
importance weight function given by
\begin{equation} \label{eq:definition-weightfunction-forward}
        \ewghtfunc{k + 1}(x, x' \parind) \eqdef \frac{g_{k + 1}(x' \parind)}
        {\adjfunc{k}{k}{x \parind}} \frac{\Mdens(x, x' \parind)}{\propdens{k}(x, x' \parind)} \eqsp.
\end{equation}
Finally, since the original target distribution
is the marginal of $\Phi^N_{k + 1}$ \wrt\
the particle position, a weighted sample approximating the former is
obtained by discarding the indices $\ind{k + 1}{i}$ and returning
$(\epart{k + 1}{i}, \ewght{k + 1}{i})_{i = 1}^N$.

The scheme is initialised by drawing $(\epart{0}{i})_{i = 1}^N$
independently from some initial instrumental distribution
$\XinitIS{0}$ on $(\Xset, \Xsigma)$ and assigning each of these
initial particles the importance weight
$\ewght{0}{i} \eqdef \ewghtfunc{0}(\epart{0}{i} \parind)$ where,
for $x \in \Xset$,
$\ewghtfunc{0}(x \parind) \eqdef g_0(x \parind) \rmd \Xinit /
\rmd \XinitIS{0}(x \parind)$.

Under suitable conditions the approximation $\filt[hat]{k}$ is
is consistent in the sense that, as $N$ tends to infinity,
$$
    \filt[hat]{k}(h \parind) \plim \filt{k}(h \parind) \eqsp,
$$
for all $\filt{k}$-integrable target functions $h$
\citep[see][for some convergence results on the auxiliary
particle filter]{douc:moulines:olsson:2008}.
In addition, as a by-product, an asymptotically consistent estimate
of the normalising constant $Z_n$ can be obtained as
\begin{equation} \label{eq:def:part:lhd:approx}
    Z^N_n \eqdef \frac{1}{N^{n + 1}}
    \left( \prod_{k = 0}^{n - 1} \sum_{\ell = 1}^N \ewght{k}{\ell}
    \adj{k}{\ell} \right) \sum_{\ell = 1}^N \ewght{n}{\ell} \eqsp.
\end{equation}

We remark that as a \textit{particle} one may view not only the
actual position $\epart{k}{i}$ at time $k$, but also the whole
trajectory $(\epart{0}{\genind{0}{i}},\epart{1}{\genind{1}{i}},\ldots, \epart{k - 1}{\genind{k - 1}{i}}, \epart{k}{i})$, where the indices $(\genind{k}{i})_{k = 0}^{n - 1}$ of the genealogical path are defined recursively backwards through $\genind{k - 1}{i} = \ind{k}{\genind{k}{i}}$ with $\genind{n}{i} = i$, of positions that led up to this current position. The particle filter may thus be used not only to approximate the filter distribution
$\filt{k}$, but also to approximate the joint smoothing distribution
$\post{0:k}{k}$ by viewing the trajectory associated
with $\epart{k}{i}$ as a draw from this distribution. The set of
all such histories is often referred to as the \textit{genealogical tree}.
The problem with this approach, in its basic form, is
that for time-points $k$ smaller than $n$, the
particles $(\epart{n}{i})_{i=1}^N$ will tend to originate from the
set small set of ancestors at time $k$. This problem is known as
\textit{degeneration} of the genealogical tree, and typically
it happens that for $k$ small enough, all particles alive at $n$
originate from the same single particle at time $k$.
The conclusion is that drawing particle trajectories ending with
$\epart{n}{i}$ will thus produce a poor estimate of the
smoothing distribution $\post{k:k}{n}$ for $k$ just a bit smaller
than $n$, as there are in practice only a small collection of
particles being sampled at that time-point. Backward sampling, to be described in the following, is a remedy to avoid this problem.

Given a sequence $(\filt[hat]{k})_{k = 0}^n$ of filter approximations obtained in a prefatory pass with the auxiliary particle filter, a particle approximation of $\post{0:n}{n}$ may, as mentioned in Section~\ref{section:prel}, be obtained by replacing each filter distribution $\filt{k}$ in \eqref{eq:smoothing:backw:decomposition} by the corresponding particle estimate $\filt[hat]{k}$. This yields the estimator
\begin{equation} \label{eq:smoothing:backw:decomposition:particle}
    \post[hat]{0:n}{n}(A \parind) \eqdef \idotsint \indic_A(\chunk{x}{0}{n}) \,
    \filt[hat]{n}(\rmd x_n \parind) \,
    \prod_{k = 0}^{n - 1} \revM{\filt[hat]{k}}(x_{k + 1}, \rmd x_k \parind)
\end{equation}
for $A \in \Xsigma^{\otimes (n + 1)}$. The estimator
\eqref{eq:smoothing:backw:decomposition:particle} was recently analysed 
in \cite{douc:garivier:moulines:olsson:2009},
establishing
its convergence
to $\post{0:n}{n}$ in several probabilistic senses. By definition \eqref{eq:backward:kernel:density:form}, each
measure $\revM{\filt[hat]{k}}(x, \cdot \parind)$, $x \in \Xset$, has
support at the particles $(\epart{k}{i})_{i = 1}^N$ only and the weight of each support point
$\epart{k}{i}$ is given by $\ewght{k}{i} \Mdens(\epart{k}{i}, x \parind) / \sum_{\ell = 1}^N \ewght{k}{\ell} \Mdens(\epart{k}{\ell}, x \parind)$. Thus the estimator $\post[hat]{0:n}{n}$ is impractical since the cardinality of its support grows exponentially with $n$. However, a \emph{draw} from $\post[hat]{0:n}{n}$ is straightforwardly obtained using the following algorithm.
\bigskip
\begin{algorithm}{alg:backward:sampler}
{\qcomment{particle-based FFBS} \label{alg:particle:FFBS}}
    run the particle filter to obtain $(\epart{k}{i}, \ewght{k}{i})_{1 \leq i \leq N, 0 \leq k \leq n}$ \\
    simulate $\backind{n} \sim (\ewght{n}{i})_{i = 1}^N$ \\
    set $\partdraw{n}{n}$ \qlet $\epart{n}{\backind{n}}$ \\
    \qfor $k \qlet n - 1$ \qto $0$ \\
    \qdo simulate $\backind{k} \sim (\ewght{k}{i} \Mdens(\epart{k}{i}, \partdraw{k + 1}{n} \parind))_{i = 1}^N$ \\
    set $\partdraw{k}{n}$ \qlet $\epart{n}{\backind{n}}$
    \qend \\
    set $\partdraw{0:n}{n}$ \qlet $(\partdraw{0}{n}, \ldots, \partdraw{n}{n})$ \\
    \qreturn $\partdraw{0:n}{n}$
\end{algorithm}
\medskip
We note, for reasons that will be clear in the coming section, 
that Algorithm~\ref{alg:particle:FFBS} provides, as a by-product of 
the forward filtering pass in Step~1, an estimate $Z_n^N$ 
(given in \eqref{eq:def:part:lhd:approx}) of the normalising constant 
$Z_n$. Since computing the normalising constant of the probability 
distribution in Step~4 involves summing over $N$ terms, the overall 
cost of executing Steps~2--7 (\ie\ the backward simulation pass) is 
$\mathcal{O}(n N)$. As noted 
in \cite{douc:garivier:moulines:olsson:2009}, 
this cost can be reduced significantly in the case where the transition 
density $\Mdens$ is bounded by some finite constant $\ub$, 
\ie\ $\Mdens(x, x') \leq \ub$ for all $(x, x') \in \Xset^2$, 
which is the case for a large class of models (\eg\ all non-linear models 
with additive Gaussian noise). Indeed, by applying instead a standard 
accept-reject scheme where a candidate $\backind{k}^\ast$ is sampled 
from the probability distribution induced by the particle weights 
$(\ewght{k}{i})_{i = 1}^N$ (whose normalising constant is obtained as a 
by-product of the forward filtering pass) and accepted with probability 
$\Mdens(\epart{k}{\backind{k}^\ast}, \partdraw{k + 1}{n} \parind) / \ub$, 
the corresponding complexity can be reduced to $\mathcal{O}(n)$. 
More specifically, 
\cite[Proposition~1]{douc:garivier:moulines:olsson:2009} 
proves
that the number of simulations per index needed for obtaining, at any time 
step $k$, $N$ indices $\backind{k}$ of $N$ conditionally independent 
replicates of the backward index chain tends to a constant in probability. 
We will apply this strategy for the implementation in 
Section~\ref{section:example}. 

%% file: partsmoothMH_PIMH.tex
Since the density of the smoothing distribution is known up to a
normalising constant, state-space models can be perfectly cast into
the framework of the Metropolis-Hastings algorithm. When applied to
smoothing in state-space models, the output of the M-H algorithm is a
Markov chain $(\MHch{0:n}{\ell})_{\ell \geq 0}$ on $\Xset^{n + 1}$ with
the following dynamics. Given $\MHch{0:n}{\ell}$, a candidate
$\chunk{X}{0}{n}^\ast$ for $\MHch{0:n}{\ell + 1}$ is produced by simulation
according to $\chunk{X}{0}{n}^\ast \sim \MHprop{n}(\MHch{0:n}{\ell}, \cdot)$,
where $\MHprop{n}$ is some proposal kernel on $\Xset^{n + 1}$; after this,
one sets
\begin{equation} \label{eq:def:standard:update}
    \MHch{0:n}{\ell + 1} =
    \begin{cases}
        \chunk{X}{0}{n}^\ast \quad \mbox{w.pr.\ }
   \ap{n}(\MHch{0:n}{\ell}, \chunk{X}{0}{n}^\ast)
    = 1 \wedge \left( \frac{\post{0:n}{n}(\chunk{X}{0}{n}^\ast)
    \MHprop{n}(\chunk{X}{0}{n}^\ast, \MHch{0:n}{\ell})}
    {\post{0:n}{n}(\MHch{0:n}{\ell}) \MHprop{n}(\MHch{0:n}{\ell},
       \chunk{X}{0}{n}^\ast)} \right) \eqsp, \\
        \MHch{0:n}{\ell} \quad \mbox{otherwise}.
    \end{cases}
\end{equation}
The initial trajectory $\MHch{0:n}{0}$ may be chosen arbitrarily.
The M-H algorithm above admits $\post{0:n}{n}$ as stationary distribution,
and under weak additional assumptions (such as Harris recurrence),
$(\MHch{0:n}{\ell})_{\ell \geq 0}$ converges in distribution to
$\post{0:n}{n}$ \citep[see \eg][for details]{roberts:rosenthal:2004}.
In order to obtain an acceptance rate close to one, one should aim
to simulate the candidates from a proposal distribution that is as
close to $\post{0:n}{n}$ as possible. Recalling
Algorithm~\ref{alg:particle:FFBS}, a natural strategy is thus to generate
the candidate using the particle-based FFBS. Indeed,
with $\mathcal{L}^N(\partdraw{0:n}{n})$ denoting the law of the draw
$\partdraw{0:n}{n}$ returned by Algorithm~\ref{alg:particle:FFBS},
\cite[Theorem~1]{olsson:ryden:stjernqvist:2010}
shows
that under rather weak assumptions there exists a constant
$C_n < \infty$ such that for all $N \geq 1$,
\begin{equation*}
    \left\| \mathcal{L}^N(\partdraw{0:n}{n})
    - \post{0:n}{n} \right\|_{\mathrm{TV}} \leq C_n/N
\end{equation*}
where $\|\cdot\|_{\mathrm{TV}}$ denotes total variation (distance);
$\|\mu -\nu\|_{\mathrm{TV}}= \sup_{\supnorm{f}\leq 1}|\mu(f)-\nu(f)|$
for probability measures  $\mu$ and $\nu$.
Unfortunately, constructing an M-H kernel based on this proposal
distribution (which is \emph{independent} of the given $\MHch{0:n}{\ell}$)
is not possible in practice since the density of
$\mathcal{L}^N(\partdraw{0:n}{n})$ is infeasible to compute.
However, the joint density of all random variables (\ie\ all indices
and particle locations drawn in the forward pass as well as the indices
obtained in the backward pass) generating the output of the
particle-based FFBS has a simple form; thus, inspired by
\cite{andrieu:doucet:holenstein:2010}, we detour this difficulty by
sampling instead a well chosen auxiliary target distribution on
the augmented state space of all these random variables. Interestingly,
it turns out that the acceptance ratio of the resulting independent
M-H sampler, which is described in Algorithm~\ref{alg:IPMH:sampler} below,
is the same as for the standard forward-smoothing-based
algorithm particle independent M-H sampler proposed
in
\cite{andrieu:doucet:holenstein:2010}.
\bigskip
\begin{algorithm}{alg:FFBS-based:IPMH:sampler}
{\qcomment{FFBS-based IM-H sampler} \label{alg:IPMH:sampler}
    \qinput{$\MHch{0:n}{\ell}$}}
    run Algorithm~\ref{alg:particle:FFBS} to obtain $\partdraw{n}{n}$
    and $Z_n^{N, \ast}$ \\
    set $\MHch{0:n}{\ell + 1} \qlet \chunk{X}{0}{n}^\ast$
    with probability
    $$
      \ap{n}(\MHch{0:n}{\ell}, \chunk{X}{0}{n}^\ast)
    = 1 \wedge \frac{Z_n^{N, \ast}}{Z_n^N}\eqsp;
    $$
    otherwise set
    $\MHch{0:n}{\ell + 1} \qlet\MHch{0:n}{\ell}$. \\
    \qreturn $\MHch{0:n}{\ell + 1}$
\end{algorithm}
\medskip

In order to derive precisely the scheme above,
denote by $\epartset{k} \eqdef (\epart{k}{1}, \ldots, \epart{k}{N})$ and
$\indset[var]{k} \eqdef (\ind{k}{1}, \ldots, \ind{k}{N})$
the collection of all particles and indices generated by the particle
filter at time step $k \geq 0$. Then the process
$(\epartset[var]{k}, \indset[var]{k})_{k \geq 1}$ is Markovian with joint
law given by the density
\begin{equation} \label{eq:joint:part:ind:law}
    \jointpartlaw{n}(\epartset{0}, \ldots, \epartset{n}, \indset{1},
       \ldots, \indset{n} \parind)
    \eqdef \left( \prod_{\ell = 1}^N \XinitIS{0}(\epart{0}{\ell} \parind)
       \right)
    \left( \prod_{k = 1}^n \prod_{\ell = 1}^N \auxprop{k}(i_k^{\ell},
          \xi_k^{\ell} \parind) \right) \eqsp,
\end{equation}
Let again $\backind{k}$, $k = n, n - 1, \ldots, 0$ denote the time-reversed
index Markov chain of the backward smoothing pass. The joint distribution
of the particle locations $(\epartset[var]{k})_{k = 0}^n$ and indices
$(\indset[var]{k})_{k = 1}^n$ obtained in the forward filtering pass
and the indices $(\backind{k})_{k = 0}^n$ of the backward smoothing pass
is given by
\begin{eqnarray}
 \lefteqn{\auxMHprop{n}(\epartset{0}, \ldots, \epartset{n},
    \indset{1}, \ldots, \indset{n}, j_0, \ldots j_n \parind)}
    \label{eq:jointpropdens} \hspace*{5mm} \\
   & \eqdef & \jointpartlaw{n}(\epartset{0}, \ldots, \epartset{n},
       \indset{1}, \ldots, \indset{n} \parind) \times
       \frac{\ewght{n}{j_n}}{\sum_{\ell = 1}^N \ewght{n}{\ell}}
        \prod_{k = 1}^n \revM{\filt[hat]{k - 1}}(\xi_k^{j_k},
         \xi_{k - 1}^{j_{k - 1}} \parind)
    \nonumber \\
   & = & \jointpartlaw{n}(\epartset{0}, \ldots, \epartset{n},
         \indset{1}, \ldots, \indset{n} \parind) \times
      \frac{\ewght{n}{j_n}}{\sum_{\ell = 1}^N \ewght{n}{\ell}}
    \prod_{k = 1}^n \frac{\ewght{k - 1}{j_{k - 1}}
       \Mdens(\xi_{k - 1}^{j_{k - 1}}, \xi_k^{j_k} \parind)}
        {\sum_{\ell = 1}^N \ewght{k - 1}{\ell} \Mdens(\epart{k - 1}{\ell},
    \epart{k}{j_k} \parind)} \eqsp,
   \nonumber
\end{eqnarray}
where the second factor is the conditional distribution of
$(\backind{k})_{k = 0}^n$ given the particle locations and indices
obtained in the forward filtering pass. Using the density
\eqref{eq:joint:part:ind:law}, the law of a draw produced by
Algorithm~\ref{alg:particle:FFBS} can be expressed as,
for $A \in \Xsigma^{\varotimes (n + 1)}$,
\begin{multline*}
    \mathcal{L}^N(\partdraw{0:n}{n})(A)
   = \E_{\auxMHprop{n}}[\mathbbm{1}_A(\epart{0}{\backind{0}}, \ldots,
       \epart{n}{\backind{n}})] \\
  = \E_{\auxMHprop{n}}[\E_{\auxMHprop{n}}[
    \mathbbm{1}_A(\epart{0}{\backind{0}}, \ldots, \epart{n}{\backind{n}}) |
    \epartset{0}, \ldots, \epartset{n}, \indset[var]{1}, \ldots,
   \indset[var]{n}]]
   =  \E_{\jointpartlaw{n}}[ \post[hat]{0:n}{n}(A)] \eqsp.
\end{multline*}

It turns out that the distribution targeted by
Algorithm~\ref{alg:IPMH:sampler} is given by
\begin{eqnarray} \label{eq:aus:MH:target}
    \auxMHtarg{n}(\epartset{0}, \ldots, \epartset{n}, \indset{1},
    \ldots, \indset{n}, j_0, \ldots, j_n \parind)
    & \eqdef & \frac{\post{0:n}{n}(\epart{0}{j_0}, \ldots,
    \epart{n}{j_n} \parind)}{N^{n + 1}} \\
&& \hspace*{-50mm}  \nonumber
   \times \frac{\jointpartlaw{n}(\epartset{0}, \ldots, \epartset{n},
   \indset{1}, \ldots, \indset{n} \parind)}{\XinitIS{0}(\epart{0}{j_0}
    \parind)
   \prod_{k = 1}^n \auxprop{k}(i_k^{j_k}, \epart{k}{j_k} \parind)}
   \times \prod_{k = 1}^n  \frac{\ewght{k - 1}{i_k^{j_k}}
    \Mdens(\epart{k - 1}{i_k^{j_k}}, \epart{k}{j_k} \parind)}
   {\sum_{\ell = 1}^N \ewght{k - 1}{\ell} \Mdens(\epart{k - 1}{\ell},
    \epart{k}{j_k} \parind)} \eqsp,
\end{eqnarray}
and the following results, whose proofs are postponed to
the appendix, are the fundamental for the construction
of this algorithm.

\begin{thm} \label{thm:correct:marginal}
    For any particle sample size $N \geq 1$, the distribution of
    $(\epart{0}{\backind{0}}, \ldots,\epart{n}{\backind{n}})$ under
    $\auxMHtarg{n}$ is $\post{0:n}{n}$.
\end{thm}

\begin{thm} \label{eq:PIMH:update}
    For any $N \geq 1$, the update produced by Algorithm~\ref{alg:IPMH:sampler}
    is a standard M-H update with target distribution $\auxMHtarg{n}$ and
    proposal distribution $\auxMHprop{n}$.
\end{thm}

Impose the following (standard) boundedness condition on the particle importance and adjustment multiplier weight functions.

\begin{assum} \label{ass:bdd:weights}
For all $0 \leq k \leq n$, $\| \ewghtfunc{k} \|_\infty < \infty$ and $\| \vartheta_k \|_\infty < \infty$.
\end{assum}

We now have the following result.

\begin{thm} \label{thm:geom:erg}
    Let Assumption~\ref{ass:bdd:weights} hold. Then there is a
$\kappa\in[0,1)$ such that for all $\ell\geq 1$ and all
$x_{0:n}\in\Xset^{n+1}$,
    $$
      \| \mathcal{L}(\MHch{0:n}{\ell}|\MHch{0:n}{0}=x_{0:n})
      -\post{0:n}{n}\|_{\rm TV}\leq \kappa^\ell.
    $$
\end{thm}
This result relies on the fact that if the ratio of target to proposal
density, here $Z_n^{N, \ast}/Z_n$, is bounded,
the independence M-H sampler converges geometrically
\citep[cf.][p.293]{andrieu:doucet:holenstein:2010}.

Finally, by applying an Azuma-Hoeffding-type exponential inequality for
geometrically ergodic Markov chains derived recently
in
\cite{douc:moulines:olsson:vanhandel:2011}
we obtain, as a corollary of Theorem~\ref{thm:geom:erg},
the following result describing the convergence of MCMC estimates
formed by the output of Algorithm~\ref{alg:IPMH:sampler}.
\begin{cor}
Let Assumption~\ref{ass:bdd:weights} hold. Then for all $N \geq 1$ there
exists a constant $c > 0$ such that for all bounded measurable functions
$h : \Xset^{n + 1} \rightarrow \R$, all $m \geq 1$, all initial
trajectories $X_{0:n}^{(0)} = x_{0:n} \in \Xset^{n + 1}$, and all $\epsilon > 0$,
$$
\prob \left( \left| \frac{1}{m} \sum_{\ell = 0}^m
h(X_{0:n}^{(\ell)}) - \post{0:n}{n}(h) \right| \geq \epsilon \right)
\leq c \exp\left( - \frac{m}{c}
\left\{  \frac{\epsilon^2}{\| h \|^2_\infty} \wedge
\frac{\epsilon}{\| h \|_\infty} \right\} \right) \eqsp.
$$
\end{cor}

%% file: partsmoothMH_RB.tex
The results above show that the acceptance probability for a proposed
trajectory obtained by backward sampling does in fact not depend on
the current or proposed trajectories themselves but only on the
likelihood estimates, computed from the set of particles and their
importance weights, underlying the respective trajectories.
Theorem~2 in \cite{andrieu:doucet:holenstein:2010} shows
that the
same holds true when tracing a trajectory backwards from the
genealogical tree, an MCMC algorithm they referred to as the
\textit{particle independent Metropolis-Hastings} (PIMH) sampler.
Therefore we can view the M-H sampler as one that proposes and
possibly accepts \textit{sets of particles} rather than
trajectories, and from the current set of particles we may choose
to simulate a trajectory either by sampling a final state and
following its genealogy backwards, or by backward sampling.

Denoting by $X^{\rm sim}_{0:n}$ a trajectory simulated by either method
using the current set of particles, we know that
$$
\E[h(X_{0:n})\mid y_{0:n}]
= \E\{\E[h(X^{\rm sim}_{0:n})\mid
(\epartset{k},\ewghtset{k})_{0\leq k\leq n} ]
\mid y_{0:n}\},
$$
where expectations are computed under the stationary distribution
of the MCMC sampler and the notation
$(\epartset{k},\ewghtset{k})_{0\leq k\leq n}$ also
contains the ancestral history of each particle, if required.
Therefore it also holds that
$$
\E[h(X_{0:n})\mid y_{0:n}]
= \E\left\{\E\left[\left.\left. J^{-1}\sum_{j=1}^J h(X^{{\rm sim},j}_{0:n})
  \,\right|\,
(\epartset{k},\ewghtset{k})_{0\leq k\leq n} \right]
\,\right|\, y_{0:n}\right\},
$$
where now $X^{{\rm sim},j}_{0:n}$ denotes one of $J$ trajectories sampled
independently.

Moreover, we can in principle remove sampling of trajectories altogether
by letting $J\to\infty$. This is equivalent to
enumerating \textit{all} possible sampled trajectories $x_{0:n}$,
computing the probability $v_{x_{0:n}}$ say of that trajectory being
sampled, and finally computing the weighted average
$\sum v_{x_{0:n}} h(x_{0:n})$. When sampling from the genealogical tree
this is possible to do, as there are only $N$ different possible
trajectories (ending at positions $\epart{n}{i}$ for $i=1,\ldots,N$).
Replacing sampling by averaging in this way is known as
\textit{Rao-Blackwellisation}.
Section~4.6 in \cite{andrieu:doucet:holenstein:2010}
certainly
does point this out, and it also provides convergence results for
the weighted average above. For backward sampling it is generally
not possible to work with all possible trajectories, as there
are typically $N^{n+1}$ of them, but for low-dimensional distributions
of $X_{0:n}$, like that of a single $X_k$ or a pair $(X_k,X_{k+1})$,
Rao-Blackwellisation is feasible. It can then be obtained by iterating
the normalised weights at time $n$ backwards through the backward
kernels, which is equivalent to the backwards pass of the forward-backward
algorithm for HMMs. Thus we obtain the smoothing probability
$v_k^i$ say that a sampled trajectory will pass through
$\epart{k}{i}$ at time $k$, and we can compute the weighted average
$\sum_{i=1}^N v_k^i h(\epart{k}{i})$. For a pair $(X_k,X_{k+1})$,
a similar computation is possible.

Computing smoothing probabilities obviously requires more
computing time than does tracing a trajectory backwards as when
sampling from the genealogical tree. However, since the
tree will have low variability at time points away from the
final time point $n$, averaging over such points will involve
summing over just one or a few particles. Backward smoothing
does not suffer from this problem, and hence we can expect better
Rao-Blackwellisation for all time-points except for the
few last ones. A compromise is however also possible, namely to simulate
a number, say 5--25, trajectories using backward sampling
and computing the average over those.
We will now take a closer look at this approach.

Write $(\epartset{}^{(r)},\ewghtset{}^{(r)})$ for the set
$(\epartset{k},\ewghtset{k})_{0\leq k\leq n}$ of particles and weights in
the $r$-th iteration of the MCMC algorithm, and let
$X^{(r,j)}_{0:n}$, $j=1,\ldots,J$, be $J$ trajectories obtained from
this set of particles using backward sampling, simulated independently.
Assume for simplicity that we wish to estimate $\E[h(X_k)\mid y_{0:n}]$
for some $k$; the discussion here generalises with only notational
changes to functionals of one than one $X$-variable.

Running $R$ MCMC iterations, $(1/RJ)\sum_{r=1}^R\sum_{j=1}^J h(X^{(r,j)}_k)$ is our estimate of $\E[h(X_k)\mid y_{0:n}]$. To express the
variance of this estimate, consider
\begin{eqnarray*}
{\rm Var}\left( \sum_{r=1}^R\sum_{j=1}^J h(X^{(r,j)}_k)\right)
& = & \E\left[ {\rm Var}\left(\left. \sum_{r=1}^R\sum_{j=1}^J h(X^{(r,j)}_k)
   \,\right|\, (\epartset{}^{(r)},\ewghtset{}^{(r)})_{r=1}^R \right)\right] \\
& + &{\rm Var}\left[ \E\left(\left. \sum_{r=1}^R\sum_{j=1}^J h(X^{(r,j)}_k)
   \,\right|\, (\epartset{}^{(r)},\ewghtset{}^{(r)})_{r=1}^R \right)\right] \\
& = & \E\left[ \sum_{r=1}^R J {\rm Var}(h(X^{\rm sim}_k) \mid
  (\epartset{}^{(r)},\ewghtset{}^{(r)}) ) \right] \\
& + & {\rm Var}\left[ \sum_{r=1}^R J \E(h(X^{\rm sim}_k)
   \mid (\epartset{}^{(r)},\ewghtset{}^{(r)})) \right] \\
& = & RJ \left\{ \sigma^2 + J R^{-1}
{\rm Var}\left[ \sum_{r=1}^R \E(h(X^{\rm sim}_k)
   \mid (\epartset{}^{(r)},\ewghtset{}^{(r)})) \right] \right\} \\
& \approx & RJ \{ \sigma^2 + J\sigma_\infty^2\},
\end{eqnarray*}
where $X^{\rm sim}$ as above denotes a generic trajectory obtained
by backward sampling,
$\sigma^2=\E[{\rm Var}(h(X^{\rm sim}_k) \mid (\epartset{},\ewghtset{}))]$,
and $\sigma_\infty^2$ is the limit as $R\to\infty$ of the normalised
variance of the sum in the second last step. This limit,
the so-called \textit{time-average variance constant} (TAVC)
in terminology from \cite[Chapter~IV.1]{asmussen:glynn:2007},
will exist if the MCMC sampler mixes not too slowly.
Thus we can approximate the variance of our estimate as
\begin{equation}\label{eq:estvar}
{\rm Var}\left( \frac{1}{RJ}\sum_{r=1}^R\sum_{j=1}^J h(X^{(r,j)}_k) \right)
\approx \frac{1}{R}(\sigma^2/J + \sigma_\infty^2).
\end{equation}

Now assume that it takes time $\tau_{\rm PF}$ to simulate one set of
particles, and that it takes time $\tau_{\rm BS}$ to simulate one
trajectory using backward sampling. The total computational cost
for obtaining the estimate above is then
$R(\tau_{\rm PF} + J \tau_{\rm BS})$. If we have a total computation
time $\tau$ available, we can minimise the right-hand side of
(\ref{eq:estvar}) under the constraint that the total computation
time is $\tau$. Treating $J$ as a continuous variable,
one finds that the optimal value of $J$ is
$$
J_{\rm opt} = \sqrt{\frac{\sigma^2/\tau_{\rm BS}}
  {\sigma_\infty^2/\tau_{\rm PF}}}.
$$
This expression is quite intuitive; if the variability of
$h(X^{\rm sim}_k)$ within a fixed set of particles tends to be large
($\sigma^2$ is large) and sampling trajectories is quick
($\tau_{\rm BS}$ is small), then we should reduce variability by drawing
many trajectories. Likewise we should do so if variability between
sets of particles is small ($\sigma_\infty^2$ is small) and
it is time-consuming to generate new sets of particles
($\tau_{\rm PF}$ is large).

In practice neither of the parameters involved above are known,
so they need to be estimated from data and run times. In the
example below we illustrate this. 

%% file: partsmoothMH_parameter.tex
Andrieu et al.\  
\cite[Section~4.4]{andrieu:doucet:holenstein:2010} 
devised an
algorithm, referred to as the \textit{particle marginal
Metropolis-Hastings} (PMMH) update, for sampling in the case where
a model parameter is included in the MCMC sampler's state space.
We will now outline, briefly, that an entirely similar approach
is applicable when trajectories are proposed using FFBS.

Thus there is a parameter (vector) $\theta$ in some
space $\Theta$, and the transition density $q$, the emission densities 
$g_k$, and the initial distribution $\rho$ may 
all depend on $\theta$. To $\theta$ belongs a prior density 
(\wrt\ some dominating measure on $\Theta$), denoted by $\pi$.
The joint posterior density of $\theta$ and $x_{0:n}$, which we
denote by $\pi_n(\theta,x_{0:n})$, is then proportional to 
$\pi(\theta)\post{0:n}{n}(x_{0:n};\theta)$.

The MCMC algorithm uses a proposal density $\MHprop{\theta}(\cdot|\cdot)$
say for proposing new values for $\theta$, and is as follows.
\smallskip
\begin{algorithm}{alg:FFBS-based:PHHM:sampler}
{\qcomment{FFBS-based PMMH sampler} \label{alg:PHHM:sampler}
    \qinput{$\theta^{(\ell)}$ and $\MHch{0:n}{\ell}$}}
sample $\theta^\ast$ from $\MHprop{\theta}(\cdot|\theta^{(\ell)})$ \\
    run Algorithm~\ref{alg:particle:FFBS}, under the parameter
   $\theta^\ast$, to obtain $\partdraw{n}{n}$ 
    and $Z_n^{N, \ast}$ \\
    set $(\theta^{(\ell+1)},\MHch{0:n}{\ell + 1})
    \qlet (\theta^\ast,\chunk{X}{0}{n}^\ast)$
    with probability 
    $$
    1 \wedge \frac{\MHprop{\theta}(\theta^{(\ell)}|\theta^\ast)Z_n^{N, \ast}}  
       {\MHprop{\theta}(\theta^\ast|\theta^{(\ell)})Z_n^N} \eqsp;
    $$
    otherwise set
    $(\theta^{(\ell+1)},\MHch{0:n}{\ell + 1})\qlet 
    (\theta^{(\ell)},\MHch{0:n}{\ell})$ \\
    \qreturn $(\theta^{(\ell+1)},\MHch{0:n}{\ell + 1})$
\end{algorithm}
\medskip

In the same fashion an in \cite{andrieu:doucet:holenstein:2010},
one may show that on an enlarged MCMC state space emcompassing
$\theta$, $(\epartset{0}, \ldots, \epartset{n})$,
$(\indset{1},\ldots,\indset{n})$ and $(j_0, \ldots, j_n)$,
the proposal density of the sampler is
$$
\MHprop{\theta}(\theta|\theta_0) 
\auxMHprop{n}(\epartset{0}, \ldots, \epartset{n}, 
    \indset{1}, \ldots, \indset{n}, j_0, \ldots j_n;\theta)
$$
where $\theta_0$ is the current parameter and
$\auxMHprop{n}(\epartset{0}, \ldots, \epartset{n}, 
    \indset{1}, \ldots, \indset{n}, j_0, \ldots j_n ;\theta)$
is as in (\ref{eq:jointpropdens}) but with dependence on $\theta$ 
included, that the density targeted by the MCMC sampler is
proportional to
$$\pi(\theta)
\auxMHtarg{n}(\epartset{0}, \ldots, \epartset{n}, \indset{1}, 
    \ldots, \indset{n}, j_0, \ldots, j_n ;\theta)
$$
with $\auxMHtarg{n}(\epartset{0}, \ldots, \epartset{n}, \indset{1}, 
    \ldots, \indset{n}, j_0, \ldots, j_n ;\theta)$
as in (\ref{eq:aus:MH:target}) but with dependence on $\theta$
included, and that the marginal distribution of
$(\theta,\epart{0}{\backind{0}}, \ldots,\epart{n}{\backind{n}})$
under this target density is the posterior 
$\pi_n(\theta,x_{0:n})$
(cf.\ Theorem~\ref{thm:correct:marginal}).
Moreover, under standard assumptions on irreducbility, the
sequence $(\theta^{(\ell)},\MHch{0:n}{\ell})$ generated by
Algorithm~\ref{alg:PHHM:sampler} will converge in
distribution to $\pi_n(\theta,x_{0:n})$
\citep[cf.][Theorem~4.4b]{andrieu:doucet:holenstein:2010}.

Since the acceptance probability of
Algorithm~\ref{alg:PHHM:sampler} does again not depend on
the current or proposed trajectories themselves but only on the 
likelihood estimates, one can just as in 
Section~\ref{sec:RB} draw multiple trajectories
from the current set of particles, or average over all of them
using backward smoothing, to reduce the variance of sample means
that approximate posterior expectations of functionals of the
latent states. Also, again the same remark applies when
trajectories are sampled backwards from the genealogical tree.

%% file: partsmoothMH_examp.tex
In this section we illustrate the methods developed above
for a state-space model often referred to as the
\textit{growth model}, and which is a standard example in the
particle filtering literature. The model is
\begin{eqnarray}
  \label{eq:systeq}
  X_k & = & \frac{1}{2}X_{k-1} + 25\frac{X_{k-1}}{1+X_{k-1}^2}
    + 8\cos(1.2 k) + V_n, \\
  \label{eq:measeq}
  Y_k & = & \frac{1}{20}X_k^2 + W_k,
\end{eqnarray}
with $X_1\sim{\rm N}(0,\sigma_0^2)$, $V_k\sim {\rm NID}(0,\sigma_V^2)$
and $W_k\sim {\rm NID}(0,\sigma_W^2)$.
Because of the square $X_{k-1}^2$ in the measurement equation
(\ref{eq:measeq}), the filter distributions for this model are in
general bimodal.

We chose parameters $\sigma_0^2=5$, $\sigma_V^2=10$ and $\sigma_W^2=1$
\citep[an example also studied in][]{andrieu:doucet:holenstein:2010},
and $N=500$ particles. We simulated a set $y_{1:50}$ of observations,
\ie\ $n=50$, and then $R=5000$ sets of particles.
We used the \textit{bootstrap filter}, \ie\ the filter with
all adjustment multiplier weights $\adj{k}{i}=1$ and proposal
kernel equal to the system dynamics; in other words,
$\proposal{k-1}(x,\cdot)$ was the Gaussian density with mean as in
the right-hand side of (\ref{eq:systeq}) and with variance
$\sigma_W^2$. For the bootstrap filter the importance weights
$\ewght{k + 1}{i}(x,x')$ simply become the emission densities
$g_{k+1}(x')$. The number of accepted proposed sets of particles
was 1515, yielding an empirical acceptance ratio $1515/R\approx 0.30$.
We did not use a burn-in period at all, as the output showed no
signs of a significant initial transient.

With the aim of estimating $\E[X_k\mid y_{1:n}]$ for each $k$,
we did in each sweep of the MCMC algorithm, \ie\ for each
current set of particles,
\begin{itemize}
\item[(i)] simulate one trajectory by tracing the genealogical
tree backwards;
\item[(ii)] compute the Rao-Blackwellised average, for each $X_k$, over
all $N$ backward trajectories from the genealogical tree;
\item[(iii)] simulate $J=25$ trajectories using backward sampling;
\item[(iv)] run the backward smoothing algorithm to compute
a smoothed average of $X_k$, which is the same as Rao-Blackwellising
backward sampling.
\end{itemize}
We denote these four methods by \textit{GT}, \textit{GTRB},
\textit{BS} and \textit{BSM} respectively. Thus GT is what it
referred to as PIMH 
in
\cite{andrieu:doucet:holenstein:2010}.
Backward sampling was done using the importance sampling (IS) scheme
in
\cite[Algorithm~1]{douc:garivier:moulines:olsson:2009}.
This scheme avoids computing all backward transition probabilities
when extending a trajectory one step backwards, although we
did abort IS, computed all transition probabilities and used them
to simulate the state in question after 15 failed IS proposals.
The average IS acceptance rate over all sets of particles and
time-points was 16\%.

\begin{figure}[tbh]
\begin{center}
\includegraphics[width=9.0cm]{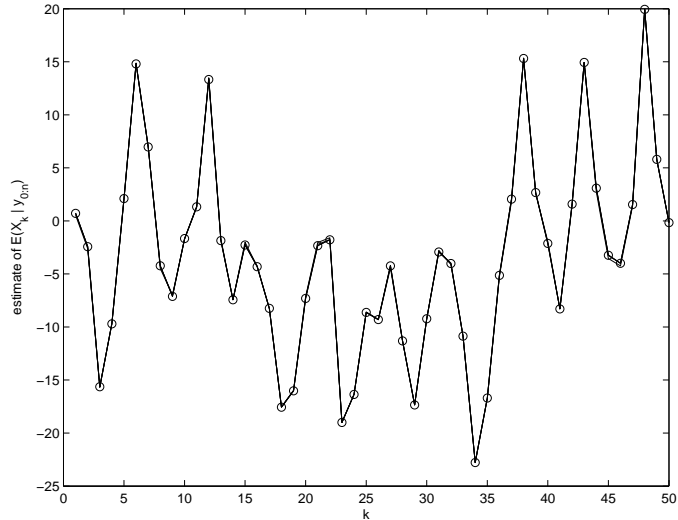}
\end{center}
\caption{Estimates of $\E[X_k\mid y_{1:n}]$ computed as sample means
from the methods GT, GTRB, BS and BSM. The four curves overlap and
are not distinguishable.}
\label{fig:means}
\end{figure}

Sample averages over the $R$ sets of particles, and, in the case
of backward sampling, over the $J$ simulated trajectories for
each set of particles, are shown in Figure~\ref{fig:means}.
Obviously all methods provide the same result, which they should,
so the differences lie in the variances.

For BSM we have the expression $\sigma_{\infty,{\rm BSM},k}^2/R$ for
the asymptotic variance, where $\sigma_{\infty,{\rm BSM},k}^2$ is as
in Section~\ref{sec:RB}; observe that the expression
$\E[h(X^{\rm sim}_k) \mid {\bf\xi}^{(r)}]$, with $h$ as the
identity function, is indeed the mean of $X_k$ obtained with
backward smoothing. Here we also include a subindex $k$ as
this variance will depend on $k$, and also a subindex
BSM as we will require similar variances for GT and GTRB.
For BS we have the asymptotic variance
$(\sigma_k^2/J + \sigma_{\infty,{\rm BSM},k}^2)/R$ as in (\ref{eq:estvar}),
where subindex $k$ in $\sigma_k^2$ again denotes dependence on
time-index $k$.
The asymptotic variances of GT and GTRB we write as
$\sigma_{\infty,{\rm GT},k}^2/R$ and $\sigma_{\infty,{\rm GTRB},k}^2/R$
respectively, where $\sigma_{\infty,{\rm GT},k}^2$ and
$\sigma_{\infty,{\rm GTRB},k}^2$ are TAVCs defined similarly as
$\sigma_{\infty,{\rm RB},k}^2$ but for one trajectory sampled from
genealogical tree and for the weighted average over all such
trajectories respectively.

\begin{figure}[tbh]
\begin{center}
\includegraphics[width=9.0cm]{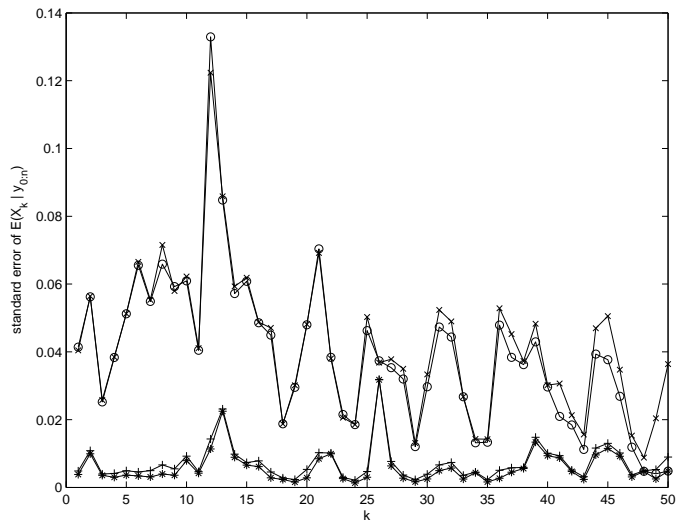}
\end{center}
\caption{Standard errors of estimates of $\E[X_k\mid y_{1:n}]$ for
the methods GT ($\times$), GTRB ($\circ$), BS ($+$) and BSM ($\ast$).}
\label{fig:stderrs}
\end{figure}

In practice neither of these variances are known, and we need to
estimate them from the simulations. We estimated $\sigma_k^2$
by first for each set of particles computing the sample variance
of all $J$ trajectories $X^{{\rm sim},j}_k$ obtained by backward
sampling, and then computing the average of these sample variances
over all $R$ sets of particles. The TAVCs $\sigma_\infty^2$ were
estimated by summing up estimated autocovariances over lags
$|\ell|<\sqrt{R}$, weighted by $(1-|\ell|/R)$
\citep[cf.][p.~59]{brockwell:davis:2002}.
Inserting these variance estimates into the expressions
for asymptotic variances and taking square roots, yield
standard errors for the respective estimates of
$\E[X_k\mid y_{1:n}]$, shown in Figure~\ref{fig:stderrs}.

We see that GT has standard errors larger than those of
BS and BSM, which is to be expected as GT samples a single
trajectory while BS samples $J=25$ trajectories and BSM averages
over all of them. We also see that the standard errors of GT and GTRB
are close to  identical expect towards the final time-point $n$.
This is a result of the degeneracy of the genealogical tree as
for $k$ just a bit less than $n$ there are only one or a few ancestors
with descendants alive at time $n$, and then Rao-Blackwellisation
(GTRB) adds little compared to just sampling (GT). For $k\geq 43$ say
GTRB however does better, and it is on par with BSM for $k\geq 48$;
for such late time-points the final particles' ancestries have not
coalesced and at time $n$ GTRB and BSM are equivalent.
Comparing BS and BSM we find that they have similar standard errors,
and this is because the term $\sigma_k^2/J$, with some exceptions $k$,
is smaller than $\sigma_{\infty,{\rm BSM},k}^2$ for $J=25$.

\begin{figure}[tbh]
\begin{center}
\includegraphics[width=9.0cm]{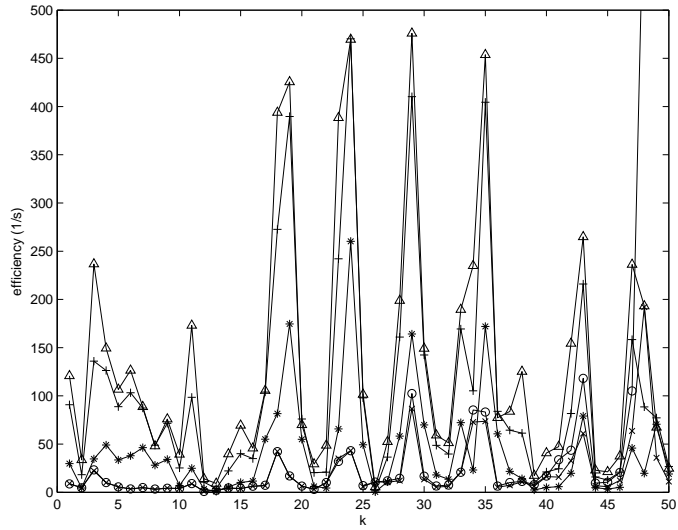}
\end{center}
\caption{Estimated efficiency for estimating $\E[X_k\mid y_{1:n}]$
for the methods GT ($\times$), GTRB ($\circ$), BS with $J=25$
trajectories ($+$), BS with $J=7$ trajectories ($\Delta$)
and BSM ($\ast$). The $y$-axis is truncated; GTRB reaches efficiencies
about 900~s$^{-1}$ for the final time points.}
\label{fig:eff}
\end{figure}

Comparing standard errors without comparing execution times does not
provide the full picture however, and for that reason we introduce
a measure of precision per computational effort, defined as
inverse variance over computation time. We refer to this measure
as \textit{efficiency}, and we can estimate it using
inverse squared standard errors over  measured computation times.
The computation time of each method was measured using the function
\verb+cputime+ in \verb+Matlab+, the software used for all simulations.
Figure~\ref{fig:eff} plots these estimates. We see that BS
is better than BSM, which in turn is better than GT and GTRB which
perform about equally. The exception is the last few time-points
for which GTRB, which is fast, does very well.
The ratios of efficiencies for BS vs.\ GT (for all $k$) ranges from 
0.19 to 30.7, with 48 (out of 50) of them being larger than one and 
their geometric mean being 5.4. For BSm vs.\ GT the corresponding figures 
are 0.04, 11.4, 36 and 1.8 respectively.

Having said that, we remark that figures like these crucially depend
on software and implementation. GTRB is fast because \verb+Matlab+
does vectorised operations quickly, and we also believe that BS has
a slight disadvantage from slow random number generation in
\verb+Matlab+. In addition the resolution of \verb+cputime+ appears
to be 10~ms, which may not be short enough to provide accurate
measures of execution times (as we measured the time of each call
to functions performing the various methods).

As discussed in Section~\ref{sec:RB}, we can choose the number $J$
of trajectories sampled in the BS method to achieve the best
variance/cost performance. This number varies quite a bit \wrt\
$k$ however, with the minimum value of $J_{\rm opt}$ (viewed as
a continuous variable) being 0.74 and the largest 18.9. As
a compromise we chose the geometric mean 6.98, rounded to $J=7$
(the arithmetic mean is 8.12). The estimated efficiency for this
$J$ is also plotted in Figure~\ref{fig:eff}, and we see that there
is improvement over $J=25$ that is mostly marginal, but
for some $k$ notable. The ratios of efficiencies for this optimised
BS vs.\ GT range from 0.50 to 37.6, with 49 (out of 50) of them
being larger than one and their geometric mean being 7.5.

%% file: partsmoothMH_proofs.tex
\subsection{Proof of Theorem~\ref{thm:correct:marginal}}

To prove the statement we simply carry through the marginalisation. Thus let $\boldsymbol{\xi}_k^{- \ell} \eqdef (\epart{k}{i})_{1 \leq i \leq N, i \neq \ell}$ and define $\indset{k}^{- \ell}$ analogously; the marginal of $\auxMHtarg{n}$ with respect to $(\epart{0}{\backind{0}}, \ldots, \epart{n}{\backind{n}}, \backind{0}, \ldots, \backind{n})$ is then obtained by integrating $\auxMHtarg{n}$ over $(\indset{k}, \boldsymbol{\xi}_k^{- j_k})_{k = 0}^n$. We start with integrating over $\indset{n}$ and $\boldsymbol{\xi}_n^{- j_n}$ according to
\begin{multline} \label{eq:first:integration}
    \auxMHtarg{n}(\epartset{0}, \ldots, \epartset{n - 1}, \epart{n}{j_n}, \indset{1}, \ldots, \indset{n - 1}, j_0, \ldots, j_n \parind)
    \\ = \sum_{\indset{n}} \int \auxMHtarg{n}(\epartset{0}, \ldots, \epartset{n}, \indset{1}, \ldots, \indset{n}, j_0, \ldots, j_n \parind) \, \rmd \boldsymbol{\xi}_n^{- j_n}
    \\ = \frac{\post{0:n}{n}(\epart{0}{j_0}, \ldots, \epart{n}{j_n} \parind)}{N^{n + 1}} \times \left( \prod_{k = 1}^{n - 1}  \frac{\ewght{k - 1}{i_k^{j_k}} \Mdens(\epart{k - 1}{i_k^{j_k}}, \epart{k}{j_k} \parind)}{\sum_{\ell = 1}^N \ewght{k - 1}{\ell} \Mdens(\epart{k - 1}{\ell}, \epart{k}{j_k} \parind)} \right)
    \\ \times \left( \sum_{\mathbf{i}_n^{- j_n}} \int \frac{\jointpartlaw{n}(\epartset{0}, \ldots, \epartset{n}, \indset{1}, \ldots, \indset{n} \parind)}{\XinitIS{0}(\epart{0}{j_0} \parind) \prod_{k = 1}^n \auxprop{k}(\ind{k}{j_k}, \epart{k}{j_k} \parind)} \, \rmd \boldsymbol{\xi}_n^{- j_n} \right)
    \\ \times \left( \sum_{i_n^{j_n}} \frac{\ewght{n - 1}{i_n^{j_n}} \Mdens(\epart{n - 1}{i_n^{j_n}}, \epart{n}{j_n} \parind)}{\sum_{\ell = 1}^N \ewght{n - 1}{\ell} \Mdens(\epart{n - 1}{\ell}, \epart{n}{j_n} \parind)} \right) \eqsp.
\end{multline}
In the expression above,
\begin{equation} \label{eq:sum:factor:equals:one}
    \sum_{i_n^{j_n}} \frac{\ewght{n - 1}{i_n^{j_n}} \Mdens(\epart{n - 1}{i_n^{j_n}}, \epart{n}{j_n} \parind)}{\sum_{\ell = 1}^N \ewght{n - 1}{\ell} \Mdens(\epart{n - 1}{\ell}, \epart{n}{j_n} \parind)} = 1
\end{equation}
and
\begin{multline} \label{eq:particle:density:marginalisation}
    \sum_{\mathbf{i}_n^{- j_n}} \int \frac{\jointpartlaw{n}(\epartset{0}, \ldots, \epartset{n}, \indset{1}, \ldots, \indset{n} \parind)}{\XinitIS{0}(\epart{0}{j_0} \parind) \prod_{k = 1}^n \auxprop{k}(i_k^{j_k}, \epart{k}{j_k} \parind)} \, \rmd \boldsymbol{\xi}_n^{- j_n}
    \\ = \frac{\jointpartlaw{n}(\epartset{0}, \ldots, \epartset{n - 1}, \indset{1}, \ldots, \indset{n - 1} \parind)}{\XinitIS{0}(\epart{0}{j_0} \parind) \prod_{k = 1}^{n - 1} \auxprop{k}(i_k^{j_k}, \epart{k}{j_k} \parind)} \eqsp.
\end{multline}
Combining \eqref{eq:first:integration}--\eqref{eq:particle:density:marginalisation} yields
\begin{multline} \label{eq:first:marginal}
    \auxMHtarg{n}(\epartset{0}, \ldots, \epartset{n - 1}, \epart{n}{j_n}, \indset{1}, \ldots, \indset{n - 1}, j_0, \ldots, j_n \parind)
    \\ = \frac{\post{0:n}{n}(\epart{0}{j_0}, \ldots, \epart{n}{j_n} \parind)}{N^{n + 1}} \times \frac{\jointpartlaw{n}(\epartset{0}, \ldots, \epartset{n - 1}, \indset{1}, \ldots, \indset{n - 1} \parind)}{\XinitIS{0}(\epart{0}{j_0} \parind) \prod_{k = 1}^{n - 1} \auxprop{k}(i_k^{j_k}, \epart{k}{j_k} \parind)}
    \\ \times \prod_{k = 1}^{n - 1}  \frac{\ewght{k - 1}{i_k^{j_k}} \Mdens(\epart{k - 1}{i_k^{j_k}}, \epart{k}{j_k} \parind)}{\sum_{\ell = 1}^N \ewght{k - 1}{\ell} \Mdens(\epart{k - 1}{\ell}, \epart{k}{j_k} \parind)} \eqsp.
\end{multline}
Now, by integrating \eqref{eq:first:marginal} with respect to $(\indset{n - 1}, \boldsymbol{\xi}_{n - 1}^{- j_{n - 1}})$ and repeating the same procedure for $(\indset{n - 2}, \boldsymbol{\xi}_{n - 2}^{- j_{n - 2}}), \ldots, (\indset{1}, \boldsymbol{\xi}_1^{- j_1})$ and finally $\boldsymbol{\xi}_0^{- j_0}$ we obtain the marginal density
\begin{equation} \label{eq:marg:path:indices}
    \auxMHtarg{n}(\epart{0}{j_0}, \ldots, \epart{n}{j_n}, j_0, \ldots, j_n \parind) = \frac{\post{0:n}{n}(\epart{0}{j_0}, \ldots, \epart{n}{j_n} \parind)}{N^{n + 1}} \eqsp.
\end{equation}
Finally, for any rectangle $A = A_0 \times A_1 \times \cdots \times A_n$ in $\Xsigma^{\varotimes (n + 1)}$,
\begin{multline*}
\prob_{\auxMHtarg{n}} \left( \epart{0}{J_0} \in A_0, \ldots, \epart{n}{J_n} \in A_n \right) \\ = \sum_{j_{0:n}} \int_{A_0} \cdots \int_{A_n} \auxMHtarg{n}(\epart{0}{j_0}, \ldots, \epart{n}{j_n}, j_0, \ldots, j_n \parind) \, \rmd \epart{0}{j_0} \cdots \rmd \epart{n}{j_n} = \post{0:n}{n}(A) \eqsp,
\end{multline*}
implying that these measures are identical on $\Xsigma^{\varotimes (n + 1)}$.
We complete the proof by noting that the arguments above apply independently of the particle sample size $N$.

\subsection{Proof of Theorem~\ref{eq:PIMH:update}}

It is enough to prove that
$$
    \mathcal{R} \eqdef \frac{\auxMHtarg{n}(\epartset{0}, \ldots, \epartset{n}, \indset[var]{1}, \ldots, \indset[var]{n}, J_0, \ldots, J_n \parind)}{\auxMHprop{n}(\epartset{0}, \ldots, \epartset{n}, \indset[var]{1}, \ldots, \indset[var]{n}, J_0, \ldots J_n \parind)} = \frac{Z_n^N}{Z_n} \eqsp,
$$
where $Z_n^N$ is defined in \eqref{eq:def:part:lhd:approx}. Using that
$$
    \auxprop{k}(\ind{k}{\backind{k}}, \epart{k}{\backind{k}} \parind) = \frac{\ewght{k - 1}{\ind{k}{\backind{k}}}
    \adj{k - 1}{\ind{k}{\backind{k}}}}{\sum_{\ell = 1}^N \ewght{k - 1}{\ell}
    \adj{k - 1}{\ell}}
    \propdens{k - 1}(\epart{k - 1}{\ind{k}{\backind{k}}}, \epart{k}{\backind{k}} \parind) \eqsp,
$$
we obtain
\begin{multline} \label{eq:MH:ratio-1}
    \mathcal{R} = \frac{\post{0:n}{n}(\epart{0}{\backind{0}}, \ldots, \epart{n}{\backind{n}} \parind) \prod_{k = 1}^n \{ \Mdens(\epart{k - 1}{\ind{k}{\backind{k}}}, \epart{k}{\backind{k}} \parind) \sum_{\ell = 1}^N \ewght{k - 1}{\ell} \adj{k - 1}{\ell} \} \sum_{\ell = 1}^N \ewght{n}{\ell}}{N^{n + 1} \XinitIS{0}(\epart{0}{\backind{0}} \parind) \ewght{0}{\backind{0}} \prod_{k = 1}^n \{ \ewght{k}{\backind{k}} \Mdens(\epart{k - 1}{\backind{k - 1}}, \epart{k}{\backind{k}} \parind) \adj{k - 1}{\ind{k}{\backind{k}}} \propdens{k - 1}(\epart{k - 1}{\ind{k}{\backind{k}}}, \epart{k}{\backind{k}} \parind) \} }\eqsp.
\end{multline}
Now, by the definition \eqref{eq:definition-weightfunction-forward} of the importance weights,
\begin{equation} \label{eq:weight:prod:id}
\ewght{k}{\backind{k}} \adj{k - 1}{\ind{k}{\backind{k}}} \propdens{k - 1}(\epart{k - 1}{\ind{k}{\backind{k}}}, \epart{k}{\backind{k}} \parind) = g_k(\epart{k}{\backind{k}} \parind) \Mdens(\epart{k - 1}{\ind{k}{\backind{k}}}, \epart{k}{\backind{k}} \parind)
\end{equation}
and plugging the identity \eqref{eq:weight:prod:id} into \eqref{eq:MH:ratio-1} gives, using that, in addition, $\XinitIS{0}(\epart{0}{\backind{0}} \parind) \ewght{0}{\backind{0}} = \Xinit (\epart{0}{\backind{0}} \parind) g_0(\epart{0}{\backind{0}} \parind)$,
\begin{equation*}
    \mathcal{R} = \frac{\post{0:n}{n}(\epart{0}{\backind{0}}, \ldots, \epart{n}{\backind{n}} \parind) \sum_{\ell = 1}^N \ewght{n}{\ell} \prod_{k = 1}^n \sum_{\ell = 1}^N \ewght{k - 1}{\ell} \adj{k - 1}{\ell}}{N^{n + 1} \Xinit(\epart{0}{\backind{0}} \parind) g_0(\epart{0}{\backind{0}} \parind) \prod_{k = 1}^n \{ g_k(\epart{k}{\backind{k}} \parind) \Mdens(\epart{k - 1}{\backind{k - 1}}, \epart{k}{\backind{k}} \parind) \}} \eqsp.
\end{equation*}
Finally, we conclude the proof by noting that
$$
    \post{0:n}{n}(\epart{0}{\backind{0}}, \ldots, \epart{n}{\backind{n}} \parind) Z_n = \Xinit(\epart{0}{\backind{0}} \parind) g_0(\epart{0}{\backind{0}} \parind) \prod_{k = 1}^n \{ g_k(\epart{k}{\backind{k}} \parind) \Mdens(\epart{k - 1}{\backind{k - 1}}, \epart{k}{\backind{k}} \parind) \} \eqsp.
$$

